\documentclass[10pt,conference]{IEEEtran}
\IEEEoverridecommandlockouts

\usepackage[
    backend=biber,
    maxcitenames=2,
    minbibnames=3,
    mincitenames=1,
    maxbibnames=3,
    uniquelist=false,
    style=ieee,
    url=false,
    doi=true,
    eprint=true,
    isbn=false,
    ]{biblatex}
\newbibmacro*{title+link}[1]{%
  \iffieldundef{doi}
    {%
      \iffieldundef{url}
        {#1}%
        {\href{\thefield{url}}{\textcolor{teal}{#1}}}%
    }
    {\href{https://doi.org/\thefield{doi}}%
       {\textcolor{teal}{#1}}}%
}

\DeclareFieldFormat
  [online,article,inbook,incollection,inproceedings,patent,thesis,unpublished,misc]
  {title}{%
    \usebibmacro{title+link}{#1}%
  }

\renewbibmacro*{doi+eprint+url}{}
\renewbibmacro*{url+urldate}{}

\DeclareFieldFormat{pages}{}
\AtEveryBibitem{
	\clearfield{pages}
	\clearfield{isbn}
	\clearfield{eprint}
	\clearfield{issn}
	\clearfield{abstract}
	\clearfield{volume}
	\clearfield{number}
	\clearfield{month}
}
\usepackage{amsmath,amssymb,amsfonts}
\usepackage{graphicx}
\usepackage{xcolor}

\usepackage{tabularray}
\UseTblrLibrary{booktabs}
\usepackage[inline]{enumitem}
\usepackage[colorlinks=true,urlcolor=teal,
            linkcolor=teal,citecolor=teal,
            pdftitle={Works on My QPU: Reproducibility in Quantum Computing Research},
            pdfauthor={Köster, Franz, Zec, Hoess, Ramsauer, Mauerer}]{hyperref}
\usepackage{glossaries-extra}
\usepackage{tikz}
\usetikzlibrary{positioning,shadows,shapes.geometric,decorations.pathreplacing,calc,arrows.meta}
\usepackage{csquotes}
\usepackage{xspace}
\usepackage{orcidlink}
\usepackage{subcaption}

\usepackage{tabularx}
\usepackage{multirow}
\usepackage{booktabs}

\setabbreviationstyle[acronym]{long-short}
\glssetcategoryattribute{acronym}{nohyperfirst}{true}

\newacronym{nisq}{NISQ}{noisy intermediate-scale quantum}
\newacronym{qaoa}{QAOA}{quantum approximate optimisation algorithm}
\newacronym{ftqc}{FTQC}{fault tolerant quantum computer}
\newacronym{qc}{QC}{quantum computing}
\newacronym{qml}{QML}{quantum machine learning}
\newacronym{vqe}{VQE}{variational quantum eigensolver}
\newacronym{ml}{ML}{machine learning}
\newacronym{qpu}{QPU}{quantum processing unit}
\newacronym{llm}{LLM}{large language model}

\newcommand{\eg}{\emph{e.g.}\xspace}
\newcommand{\ie}{\emph{i.e.}\xspace}

\newcommand{\partemail}[2]{\href{mailto:#1@#2}{#1}}

\definecolor{lfd1}{HTML}{000000}
\definecolor{lfd2}{HTML}{E69F00}
\definecolor{lfd3}{HTML}{999999}
\definecolor{lfd4}{HTML}{009371}
\definecolor{lfd5}{HTML}{BEAED4}
\definecolor{lfd6}{HTML}{ED665A}
\definecolor{lfd7}{HTML}{1F78B4}

\setlist[enumerate]{label=({\arabic*})}

\begin{document}

\title{Works on my QPU: \\ Reproducibility in Quantum Computing Research}
\author{
    \IEEEauthorblockN{%
        Dominik Köster\IEEEauthorrefmark{1}\orcidlink{0009-0001-0230-3123},
        Maja Franz\IEEEauthorrefmark{1}\orcidlink{0000-0002-2801-7192},
        Benjamin Zec\IEEEauthorrefmark{1}\orcidlink{0009-0009-9865-8217},
        Nicole Hoess\IEEEauthorrefmark{1}\orcidlink{0009-0002-6194-5887},
        Ralf Ramsauer\IEEEauthorrefmark{1}\orcidlink{0009-0007-4033-8515},
        Wolfgang Mauerer\IEEEauthorrefmark{1}\IEEEauthorrefmark{2}\orcidlink{0000-0002-9765-8313}%
    }%
    \IEEEauthorblockA{%
        \IEEEauthorrefmark{1}%
        Technical University of Applied Sciences Regensburg, Germany,\\%
        \{%
	\partemail{dominik.koester}{othr.de},
        \partemail{maja.franz}{othr.de},
        \partemail{benjamin.zec}{othr.de},
        \partemail{nicole.hoess}{othr.de},
	\partemail{ralf.ramsauer}{othr.de},
	\partemail{wolfgang.mauerer}{othr.de}\}@othr.de%
    }%
    \IEEEauthorblockA{%
        \IEEEauthorrefmark{2}%
        Siemens AG, Foundational Technology, Munich, Germany%
    }%
}%

\maketitle

\begin{abstract}
Quantum computing research increasingly depends on complex software stacks,
yet the reproducibility of published results does not receive the priority and longevity mandated by recommendations of large international scientific bodies and best practices in software-centric systems research.
In this paper, we present a combined manual and automated large-scale analysis of the reproducibility landscape in quantum computing research, quantify shortcomings, and derive actionable steps forward. 

We manually evaluate a curated sample of 127 papers using a five-question framework that covers code availability, environment specification, documentation, hardware description, and executability.
To place these findings in a broader context, we conduct an automated large-scale screening of nearly 5000 quantum computing papers for the same reproducibility indicators.
Our manual analysis reveals that only 24.4\% of the sampled papers provide code artefacts, and among those,
64.5\% fail to execute successfully in a clean environment.
This assessment is corroborated by a large-scale automated analysis that yields a consistent code availability rate of 26.8\%.
Further, it shows that approximately one-third of the papers with accessible code lack machine-readable environment specifications.
  
The results in this paper indicate that reproducibility is not yet consistently achieved in quantum computing research.
In response, we outline a set of practical recommendations that address the observed failure modes and illustrate how reproducibility can be improved in practice.
\end{abstract}

Reproducibility is an accepted backbone of scientific progress, yet continues to pose challenges in many dimensions~\cite{gonzalez-barahona_reproducibility_2012, mauerer:22:icse}.
Ensuring reproducible research artefacts remains challenging across
disciplines~\cite{trautsch_addressing_2018, gonzalez-barahona_revisiting_2023, liang_can_2024}.
In \gls{qc}, this challenge is dominated by the
rapid dynamics of development processes, evolving tooling, and experimental
hardware infrastructures~\cite{tao:23:architecture} that are often not designed for long-term stability
and availability, including cloud-based systems~\cite{senapati_towards_2023, carbonelli:24:scenario}.
As \gls{qc} research relies on complex software environments, the
reproducibility of experimental results depends not only on algorithmic
descriptions~\cite{munasinghe_knowledge-based_2025}, but also on the
availability and precise configuration of software
dependencies.

A natural assumption is that reproducibility issues primarily arise from
environment drift over time, such as outdated dependencies or deprecated APIs.
However, during our reproduction attempts of publicly available \gls{qc} artefacts,
we observed that many projects failed to execute even in freshly provisioned
environments shortly after publication. These failures occurred independently
of long-term dependency changes, and suggest that some artefacts may never have
been fully reproducible outside the original development setup. This
observation points to a broader issue: implicit assumptions about the
development environment, such as unreferenced locally installed dependencies,
or undocumented configuration steps. These remain unrecorded and therefore
hinder reproducibility.

Publishing reproducibility artefacts (\eg, code) alongside research is
generally highly commendable and a necessary precondition for reproducibility
in \gls{qc}~\cite{mauerer_1-2-3_2022}. In this work, we conduct an empirical study of a filtered sample drawn from a cross-section of two scientific databases. For papers within this sample that provide publicly available code artefacts,
we attempt reproduction in a clean environment, following only the provided
documentation. We record reproduction outcomes and perform a qualitative analysis 
of observed failures by identifying recurring patterns that reveal implicit environmental assumptions, including 
incomplete dependency specifications, undocumented system-level requirements, 
reliance on transient container images, and hidden local state. While we observe these patterns in the underexamined field of \gls{qc}, several are not specific to it and can arise in software projects more broadly.

Our findings indicate that reproducibility challenges in \gls{qc} are not solely
caused by temporal environmental drift, but frequently stem from incomplete or
implicit environment specifications. Existing approaches such as virtual
environments, dependency files, or containerisation~\cite{mauerer:22:icse,mauerer_1-2-3_2022} mitigate parts of the
problem but often remain insufficient to capture the full execution context.
Based on our observations, we discuss declarative environment specifications as
a promising mitigation strategy. In particular, we highlight how languages and
tools such as Nix and devenv.sh enable explicit and reproducible environment
definitions~\cite{malka_reproducibility_2024} that reduce implicit assumptions, and allow for long-term
reproducibility even in the face of evolving dependencies and hardware.

In this paper, we claim the following contributions:
\begin{enumerate}
	\item A large-scale semi-automatic analysis of the availability and quality characteristics of nearly 5000 \gls{qc} papers from 2021 to 2026, which provides a quantitative estimate of the proportion of recent \gls{qc} works that include reproduction packages.
	\item The manual analysis of a subset of these \gls{qc} papers with an attempt to reproduce the claimed results, in which we identify common failure modes of reproduction packages.
	\item An actionable list of recommendations to improve the resilience against temporal environmental changes and implicit assumptions of development environments.
    \item A reproduction package that operationalises these recommendations into a reusable and extendable \href{https://github.com/lfd/qce26_claim_against_measurement}{template} for future studies.
\end{enumerate}

\section{Background and related work}
The challenge of providing reproducible research artefacts has previously 
been addressed in classical software engineering. For instance, early studies 
propose
a minimal process with related elements required for empirical studies~\cite{gonzalez-barahona_reproducibility_2012}, while
later approaches present actionable guidelines to build
self-contained Docker images. 
Beyond code repositories and virtual environments, these images
include an automated end-to-end study pipeline
with all relevant dependencies~\cite{mauerer:22:icse} 
and can be provided in long-term repositories such as Zenodo.
Despite these advances, 
research artefacts are still often not published at all, only in parts,
or as a diverse~\cite{trautsch_addressing_2018, liang_can_2024},
possibly unusable collection of scripts~\cite{gonzalez-barahona_revisiting_2023}.

In \gls{qc}, further limitations in reproducibility arise due to 
characteristics of 
software stacks. User-written quantum programs are compiled into an abstract
instruction sequence~\cite{ramsauer:25:accelerator}, 
such as a quantum circuit~\cite{younis_quantum_2022, felix:23:imperfections, krueger:25:loop}, 
defining a logical schedule~\cite{schmidbauer:26:quadratisation}.
As quantum
computers vary significantly in their physical implementation~\cite{ramsauer:25:accelerator, felix:23:imperfections}, 
this instruction sequence is then transpiled into a physical schedule for a specific target QPU~\cite{veiga_reproducible_2025} and 
brought to execution via 
control pulses~\cite{shi_resource-efficient_2020}. 
Although uncertainty is also introduced at the lower layers --- 
for instance during hardware-dependent optimisations~\cite{veiga_reproducible_2025} 
or varying noise across and 
within \gls{nisq} devices~\cite{senapati_towards_2023, dasgupta_impact_2024, thelen:24:noise, maschek:25:noise} ---
this study focuses on reproducibility from a
software perspective.

The predominant programming language of popular quantum software 
frameworks such as Qiskit~\cite{javadi-abhari_quantum_2024}, 
PennyLane~\cite{bergholm_pennylane_2022}, 
Cirq~\cite{developers_cirq_2025} and 
pyQuil~\cite{smith_practical_2016} is Python.
While offering convenience~\cite{schulz_accelerating_2022} 
for user-written programs, 
the Python ecosystem is short-lived with rapid
deprecation cycles and limited backward compatibility~\cite{zhong_empirical_2025}.
Typical issues such as version conflicts, long dependency
chains~\cite{bose_autopydep_2025},
and dependency drift, referring to unexpected changes in system 
behaviour due to updates, therefore also apply to 
quantum software~\cite{fernandez-osuna_exploring_2025}. 
For instance, studies report that migrating to a new \gls{qc}
framework version caused
incompatibilities, faults~\cite{kitt_morphq_2024} and 
even substantially 
different results, for instance due to hidden changes in circuit hyperparameter defaults~\cite{cardinal_migrating_2026}. 
The authors report significant efforts to identify, debug and 
resolve root causes~\cite{kitt_morphq_2024, cardinal_migrating_2026},
which may also limit the ability to integrate new research 
projects into the rapidly evolving main frameworks. 
In addition, such issues call into question the validity and 
generalisability of study results, especially when considered
in conjunction with potentially missing research artefacts and
restricted hardware access~\cite{mauerer_1-2-3_2022}.

Reproducibility should therefore receive more attention in \gls{qc} research.
First approaches propose 
knowledge graphs to document as many workflow and hardware details as 
possible~\cite{munasinghe_knowledge-based_2025} and present a 
Docker meta-container template for quantum software experiments~\cite{mauerer_1-2-3_2022}.
While such self-contained images ensure study reproducibility 
over time, they may not be sufficient when the objective is to 
\emph{modify} or build an environment from scratch.
The Nix package manager allows for building environments from source~\cite{malka_does_2025} and has
previously been proposed for classical high-performance computing (HPC)
settings to
automate workflows and ensure reproducibility 
across environments and systems~\cite{devresse_nix_2015}. 
While recent studies demonstrate its effectiveness 
on classical systems~\cite{malka_reproducibility_2024, malka_does_2025}, 
its adoption in the context of \gls{qc} has not yet been explored.

\section{Quantitative Reproduction Study}

\begin{table}[t]
  \centering
  \caption{Reproducibility Assessment Framework for \gls{qc} Papers.}
  \label{tab:reproducibility_framework}
  \begin{tblr}{width=\linewidth,
    colspec={lX[l,2cm]X[l]},
               row{2-Z}={belowsep=0.15em}}
  \toprule
  \textbf{ID} & \textbf{Criterion} & \textbf{Questions to Answer} \\
  \midrule
  RQ1 & Code Availability & Is source code published? Where (GitHub, Zenodo, supplementary)?
       Complete reproduction package or only snippets? Runnable or illustrative only? \\
  RQ2 & Environment & Is the environment specified? Framework mentioned? Are \texttt{requirements.txt}, \texttt{Dockerfile}, or
  \texttt{environment.yml} available? \\
  RQ3 & Build \& Run Documentation & Are instructions provided? Are logs and outputs self-explanatory?\\
  RQ4 & Hardware Specification & Are hardware specifications available? Calibration snapshot? \\
  RQ5 & Executability & Does the package run without errors? Are specified dependencies installable and run without conflicts? \\
  \bottomrule
\end{tblr}
\end{table}

The reproducibility of a scientific paper in \gls{qc} typically involves computational artefacts that cannot be perfectly captured by a textual description alone.
For this reason, we expect a reproduction package to be provided alongside these artefacts.
In this section, we describe the setup and results of a quantitative analysis of the reproducibility landscape in current \gls{qc} research.
Notably, our pipeline for the reproducibility assessment is itself reproducible, with the code provided on GitHub\footnote{\url{https://github.com/lfd/qce26_repro_analysis}} and a long-term Zenodo archive\footnote{\url{https://doi.org/10.5281/zenodo.21539924}}.

\subsection{Reproducibility Criteria}
A reproduction package typically includes
documentation on the structure and use of the package, as well as a clear and complete definition of the environment needed to generate these artefacts.

Reproducing quantum-specific artefacts may additionally require execution on actual quantum hardware,
which cannot be replicated by a reproduction package per se due to associated costs and limited availability.
In this case, at least the execution logs as well as hardware specifications, such as calibration snapshots, should be provided.
Given these prerequisites, we ask whether the package is executable and the artefact can be recreated.
These reproducibility criteria lead to the research questions summarised in \autoref{tab:reproducibility_framework}, which we aim to answer for a selection of \gls{qc} research papers in this section.

\subsection{Method}

\begin{figure*}
  \centering
  \newlength{\WIDTH}\newlength{\HEIGHT}
\setlength{\WIDTH}{\linewidth}
\setlength{\HEIGHT}{4cm}

\pgfdeclarelayer{bg}
\pgfdeclarelayer{bg1}
\pgfsetlayers{bg,bg1,main}

\def\titleheight{5mm}
\def\boxwidth{1.3cm}
\def\boxheight{1.2cm}
\def\boxdist{3.8mm}
\def\paperheight{0.82cm}
\def\paperwidth{0.707*\paperheight}
\def\actorheight{1.2cm}
\def\actorwidth{1.4cm}
\def\filterwidth{1.1cm}
\def\lw{0.5pt}
\def\boxshift{1pt}

\tikzset{block/.style={text=white, fill=lfd3, draw=lfd3, line width=\lw, font=\small, minimum width=\boxwidth, minimum height=\titleheight, inner sep=0pt}}
\tikzset{blocktext/.style={font=\scriptsize, text=white}}
\tikzset{cntbox/.style={fill=lfd3!20, draw=lfd3, line width=\lw, minimum width=\boxwidth, text width=\boxwidth, inner sep=0pt, font=\tiny}}
\tikzset{cntboxtext/.style={font=\tiny}}
\tikzset{actor/.style={font=\small, text=white, fill=lfd2, draw=lfd2, line width=\lw, minimum width=\actorwidth, minimum height=\actorheight, text width=\actorwidth, align=center}}
\tikzset{results/.style={font=\small, text=white, fill=lfd4, draw=lfd4, line width=\lw, minimum width=1.55*\actorwidth, minimum height=\actorheight, text width=1.55*\actorwidth, align=center}}
\tikzset{validation/.style={font=\small, text=white, fill=lfd2, draw=lfd2, line width=\lw, minimum width=1.55*\actorwidth, minimum height=\actorheight, text width=1.55*\actorwidth, align=center}}
\tikzset{arr/.style={-{Stealth}, line width=\lw}}
\tikzset{paper/.style={fill=white, draw=black, minimum height=\paperheight, minimum width=\paperwidth, line width=\lw, font=\scriptsize, inner sep=0pt, text width=\paperwidth, align=center}}
\tikzset{code/.style={fill=white, draw=black, minimum height=\paperwidth, minimum width=\paperheight, line width=\lw, font=\scriptsize, inner sep=0pt, text width=\paperwidth, align=center}}
\tikzset{codetext/.style={text=black, font=\tiny, text width=0.9\paperwidth, align=center}}
\tikzset{folderthingi/.style={fill=white, draw=black, minimum height=0.12*\paperwidth, minimum width=0.3*\paperheight, line width=\lw, font=\scriptsize, inner sep=0pt, align=center}}
\tikzset{filtertitle/.style={block, fill=lfd3, draw=lfd3, minimum width=\filterwidth}}
\tikzset{filterbox/.style={cntbox, fill=lfd3!20, draw=lfd3, minimum height=\actorheight-\titleheight, align=center, minimum width=\filterwidth, text width=\filterwidth}}

\newcommand{\paperbox}[4]{%
  \node[paper, anchor=#4] (#1) at (#3) {#2};
  \begin{pgfonlayer}{bg1}
    \node[paper, xshift=2*\boxshift, yshift=-2*\boxshift] at (#1) {};
    \node[paper, xshift=1*\boxshift, yshift=-1*\boxshift] at (#1) {};
  \end{pgfonlayer}
}

\newcommand{\codebox}[4]{%
  \node[code, anchor=#4] (#1) at (#3) {#2};
  %\node[codetext, anchor=north] at (#1.north) {\texttt{<CODE>}};
  \node[folderthingi, anchor=south west, yshift=-\lw] (#1folderthingi) at (#1.north west) {};
  \begin{pgfonlayer}{bg1}
    %\node[folderthingi, xshift=2*\boxshift, yshift=-2*\boxshift] at (#1folderthingi) {};
    %\node[folderthingi, xshift=1*\boxshift, yshift=-1*\boxshift] at (#1folderthingi) {};
    \node[code, xshift=2*\boxshift, yshift=-2*\boxshift] at (#1) {};
    \node[code, xshift=1*\boxshift, yshift=-1*\boxshift] at (#1) {};
  \end{pgfonlayer}
}

\newcommand{\filterbox}[3]{%
  \node[filtertitle, anchor=south west] (#1filter) at ($#2 + (0,0.5*\actorheight-\titleheight)$) {Filter};
  \node[filterbox, anchor=north] (#1filtercnt) at ($(#1filter.south)$) {#3};
  \coordinate (#1filtereast) at ($(#1filter.north east)!0.5!(#1filtercnt.south east)$);
  \coordinate (#1filterwest) at ($(#1filter.north west)!0.5!(#1filtercnt.south west)$);
}

\begin{tikzpicture}
  \coordinate (tl) at (0, \HEIGHT);
  \coordinate (tr) at (\WIDTH, \HEIGHT);
  \coordinate (bl) at (0, 0);
  \coordinate (br) at (\WIDTH, 0);

  %\begin{pgfonlayer}{bg}
  %  \draw[help lines, lightgray] (0,0) grid (\WIDTH, \HEIGHT);
  %\end{pgfonlayer}

  \node[block, anchor=north west] (querytitle) at (tl){Query};
  \node[cntbox, anchor=north, minimum height=\HEIGHT-\titleheight-2*\lw, inner sep=0pt] (querybox) at (querytitle.south){};
  \node[cntboxtext, text width=\HEIGHT-\titleheight, rotate=90, minimum height=\boxwidth, inner sep=0pt] (queryboxtext) at (querybox.center) {%
  \vspace*{-1em}
  \begin{verbatim}
  "quantum computing"
    AND ("algorithm" OR "software")
    ANDNOT ("survey" OR "review")
    AND submittedDate:
    [20210101 TO 20260327]
  \end{verbatim}%
  };
  \node[actor, anchor=west] (arxiv) at ($(querytitle.north east) + (\boxdist, -0.25*\HEIGHT)$) {ArXiv};
  \node[actor, anchor=west] (semsch) at ($(querytitle.north east) + (\boxdist, -0.75*\HEIGHT)$) {Semantic \\ Scholar};

  \draw[arr] (querybox.east|-arxiv.center) -- (arxiv);
  \draw[arr] (querybox.east|-semsch.center) -- (semsch);

  \paperbox{resarxiv}{3785}{$(arxiv.east) + (\boxdist, 0)$}{west}
  \paperbox{ressemsch}{1686}{$(semsch.east) + (\boxdist, 0)$}{west}

  \draw[arr] (arxiv) -- (resarxiv);
  \draw[arr] (semsch) -- (ressemsch);

  \coordinate (mid) at ($(resarxiv.east)!0.5!(ressemsch.east)$);
  \node[actor, anchor=west] (combine) at ($(mid)+(\boxdist,0)$) {Combine \& De-Duplicate};
  \draw[arr] (resarxiv.east) -- ++ (0.4*\boxdist, 0) |- ($(combine.west) + (0, 0.1*\actorheight)$);
  \draw[arr] (ressemsch.east) -- ++ (0.4*\boxdist, 0) |- ($(combine.west) - (0, 0.1*\actorheight)$);

  \paperbox{rescombined}{4966}{$(combine.east) + (\boxdist, 0)$}{west}

  \draw[arr] (combine) -- (rescombined);

  \filterbox{nisq}{(tl-|rescombined.east) + (\boxdist, -0.25*\HEIGHT)}{\enquote{NISQ} in Title/Abstract}
  \draw[arr] (rescombined.east) -- ++ (0.4*\boxdist, 0) |- (nisqfilterwest);

  \paperbox{resnisq}{648}{$(nisqfiltereast) + (\boxdist, 0)$}{west}
  \draw[arr] (nisqfiltereast) -- (resnisq);

  \filterbox{peer}{(resnisq.east) + (\boxdist, 0)}{Peer-Reviewed \& Experimental}
  \draw[arr] (resnisq) -- (peerfilterwest);

  \paperbox{respeer}{249}{$(peerfiltereast) + (\boxdist, 0)$}{west}
  \draw[arr] (peerfiltereast) -- (respeer);

  \filterbox{both}{(respeer.east) + (\boxdist, 0)}{ArXiv \& Semantic Scholar}
  \draw[arr] (respeer) -- (bothfilterwest);

  \paperbox{resboth}{127}{$(bothfiltereast) + (\boxdist, 0)$}{west}
  \draw[arr] (bothfiltereast) -- (resboth);

  \node[actor, anchor=west] (kwsearch) at ($(resboth.east) + (\boxdist, 0)$) {Keyword-Based Search};
  \draw[arr] (resboth) -- (kwsearch);

  \paperbox{reskw}{66}{$(kwsearch.east) + (\boxdist, 0)$}{west}
  \draw[arr] (kwsearch) -- (reskw);

  \node[validation, anchor=east] (manual) at (semsch-|tr) {Manual \\Code-Repository Search};
  \draw[arr] (reskw) -- (manual.north-|reskw);

  \codebox{resmanual}{31}{$(manual.west) + (-\boxdist, 0)$}{east}
  \draw[arr] (manual) -- (resmanual);

  \coordinate (mid) at ($(manual.north)!0.5!(kwsearch.south)$);
  \begin{pgfonlayer}{bg}
    %\fill[lfd4!20] (br) -- (bothfilter|-br) -- (bothfilter|-mid) -| (br) -- cycle;
  \end{pgfonlayer}

  \node[validation, anchor=west] (llmscreening) at ($(bl-|rescombined.east) + (\boxdist, 0.25*\HEIGHT)$) {Automated Code-Repository Search};
  \draw[arr] (rescombined.east) -- ++ (0.4*\boxdist, 0) |- (llmscreening);

  \codebox{resllm}{1330}{$(llmscreening.east) + (\boxdist,0)$}{west}
  \draw[arr] (llmscreening.east) -- (resllm);

  \node[results] (results) at ($(resllm)!0.5!(resmanual)$) {Reproducibility Evaluation};
  \draw[arr] (resllm) -- (results);
  \draw[arr] (resmanual) -- (results);

\end{tikzpicture}
  \caption{Papers are retrieved via a structured query from arXiv and Semantic Scholar, merged and deduplicated. Subsequently resulting papers are either filtered through multiple stages, identified via keyword-based screening and finally subjected to manual validation, or directly submitted to automated screening for code-repositories. The results for the final reproducibility evaluation are summarised in \autoref{tab:repro_results}.}
  \label{fig:pipeline}
\end{figure*}
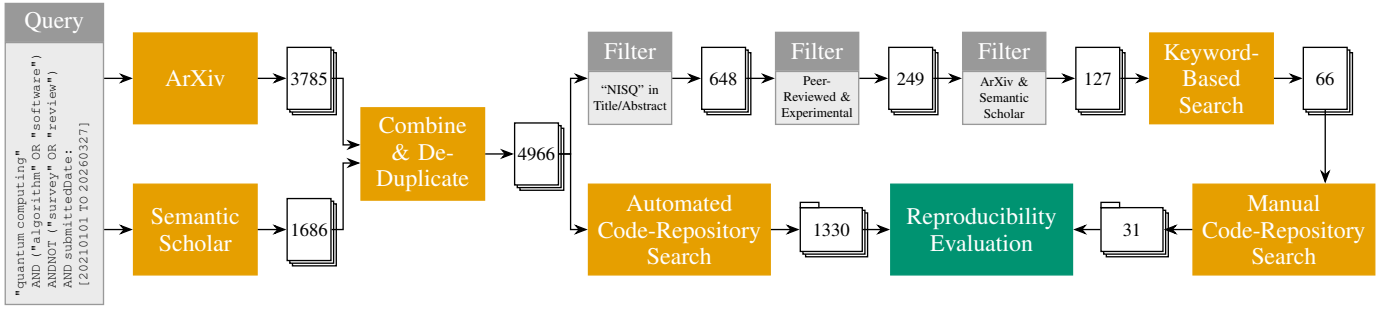

\autoref{fig:pipeline} provides an overview of the paper sampling, filtering, and manual validation pipeline, together with the corresponding quantitative results at each stage. Each of these stages is summarised in the following.

\subsubsection{Paper Sampling}
We constructed a dataset of \gls{qc} research papers by querying the databases of two scientific literature tools, \emph{arXiv}~\cite{arxiv} and \emph{Semantic Scholar}~\cite{kinney2025}.
To identify relevant publications, we formulated a structured query targeting publications in the \gls{qc} software and algorithm domain:
Specifically, we required the presence of the following terms in the title or abstract:
\begin{enumerate*}
  \item \emph{quantum computing}, and
  \item \emph{algorithm} or \emph{software}.
\end{enumerate*}
To avoid secondary literature, papers, which include the terms \emph{survey} and \emph{review} in title or abstract were explicitly excluded.
Due to our focus on the \gls{qc} research community, the search was further restricted to the arXiv category \emph{quant-ph} and to publications between
{January 1, 2021} and {March 27, 2026}.
We aggregated results from both sources based on identifiers such as DOIs, as well as title, and authorlist, yielding a unified de-duplicated paper corpus of 4966 papers.

\subsubsection{Human-validated Analysis Pipeline}
To make the size of the paper corpus manageable for manual analysis, we focus on \gls{nisq}-era research only.
Particularly, we adapt the query on the two databases to contain the keyword \emph{NISQ} in title or abstract, which results in 648 subject papers.

\paragraph{Paper Filtering}
As we are only interested in papers that contain experiments or numerical evaluations, we further filtered the paper corpus using a full-text search.
Specifically, all papers not containing the keywords that indicate code
or one of the phrases \emph{experiments}, \emph{experimental result}, or \emph{numerical} were discarded.
To increase the reliability of the sampled papers, we further restricted the dataset to
\begin{enumerate*}
  \item papers with a DOI (as a proxy for peer review), and
  \item papers that were found in both databases.
\end{enumerate*}
This step resulted in a set of 127 potentially experimental papers, which forms the starting point of the human-validated analysis pipeline.

\paragraph{Automated Screening}
\label{sec:screening}
To identify candidate papers with reproducibility artefacts, we performed another full-text scan of the available PDFs.
The scan searched for reproducibility indicators such as references to source code platform (\ie,
\begin{enumerate*}
  \item \emph{git(hub$\mid$lab$\mid$ee)},
  \item \emph{bitbucket}, or
  \item \emph{zenodo}),
\end{enumerate*}
or explicit statements of code availability (\ie,
\begin{enumerate*}
  \item \emph{source[-]code},
  \item \emph{repository},
  \item \emph{code availab(le$\mid$ility)},
  \item \emph{data availab(le$\mid$ility)},
  \item \emph{download code},
  \item \emph{reproduc(ible$\mid$tion)}, or
  \item \emph{code to reproduce}),
\end{enumerate*}
reflecting common best practices.
This automated screening resulted in 66 out of 127 papers that exhibited at least one reproducibility indicator.

\paragraph{Manual Validation}
As the automated screening favours a larger number of false positives, we performed an additional manual validation on each candidate paper from the previous step.
For the subset of confirmed papers, which do have some reproduction package linked, we further analysed the characteristics of the provided artefacts.
In particular, we assessed the artefacts that satisfy RQ1 (code availability) according to the questions RQ2 to RQ4 in \autoref{tab:reproducibility_framework}.
For those papers that satisfied both, RQ2 (environment specification) and RQ3 (documentation), we further tested for RQ5 (executability).

\subsubsection{Automated Large-Scale Analysis Pipeline}
The human-validated study provides a detailed, qualitative analysis of a sample of the \gls{qc} research landscape, which is necessarily constrained by our sampling and filtering criteria.
To contextualise these findings and assess the broader state of reproducibility across the \gls{qc} research landscape, we further conduct a larger-scale, automated analysis on the full paper corpus of 4966 papers.
For each PDF, we applied a full-text scan for the same reproducibility indicators as in the manual screening.
Additionally, we apply the following normalisation steps to reduce false positives and analyse corresponding repositories, which
\begin{enumerate*}
  \item detect repository URLs with repair heuristics for line breaks and trailing punctuation introduced by PDF text extraction,
  \item verify the accessibility of the repositories via HTTP requests, and
  \item check for environment specific files (\eg, requirements.txt, Dockerfile) and documentation indicators (\eg, README.md).
\end{enumerate*}
Based on the extracted metadata, each paper was automatically assessed against the first four questions of our framework (see \autoref{tab:reproducibility_framework}), while RQ5 was not evaluated due to the infeasibility of large-scale reproduction attempts.

\begin{table}[t]
\centering
\caption{Reproducibility indicators across the quantum computing research landscape.
\textbf{Manual analysis}: 127 considered \gls{nisq}-era papers.
Automated large-scale analysis: 4966 papers in the quantum algorithm/software domain.
Bold values correspond to the \textbf{manual analysis}.}
\label{tab:repro_results}

\begin{tabularx}{0.85\linewidth}{lrrrr}
\toprule
\textbf{Indicator} 
& \multicolumn{1}{c}{\textbf{Yes}} 
& \multicolumn{1}{c}{\textbf{Partially}} 
& \multicolumn{1}{c}{\textbf{No}} 
& \multicolumn{1}{c}{\textbf{N/A}} \\
\midrule
%\multirow{2}{*}
	{Code Availability}
  & \textbf{24.4\%} & - & \textbf{67.7\%} & \textbf{7.9\%} \\
	& 26.8\% & - & 73.2\% & - \\[0.5em]
%\midrule
%\multirow{2}{*}
	{Environment Spec.}
  & \textbf{13.4\%} & \textbf{2.4\%} & \textbf{76.3\%} & \textbf{7.9\%} \\
	& 14.6\% & 3.0\% & 82.4\% & - \\[0.5em]
%\midrule
%\multirow{2}{*}
	{Documentation}
  & \textbf{15.0\%} & \textbf{6.3\%} & \textbf{70.8\%} & \textbf{7.9\%} \\
	& 20.9\% & 1.9\% & 77.2\% & - \\[0.5em]
%\midrule
%\multirow{2}{*}
	{Hardware Spec.}
  & \textbf{2.4\%} & \textbf{1.6\%} & \textbf{69.2\%} & \textbf{26.8\%} \\
	& 6.3\% & 11.6\% & 73.2\% & 8.9\% \\[0.5em]
%\midrule
%\multirow{2}{*}
	{Executability}
  & \textbf{8.7\%} & \textbf{0.8\%} & \textbf{82.6\%} & \textbf{7.9\%} \\
  & - & - & 73.2\% & 26.8\% \\
\bottomrule
\end{tabularx}
\end{table}
\renewcommand{\arraystretch}{1}

\subsection{Results}

The results for the human-validated manual and large-scale analysis are summarised in \autoref{tab:repro_results} and described below.
Overall, both analyses paint a consistent picture of the current state of reproducibility practices in \gls{qc} research. The alignment in code availability rates between the curated manual sample (24.4\%) and the broader automated corpus (26.8\%) is not coincidental: the manual sampling filters for experimental papers with a DOI and cross-database presence, criteria that are orthogonal to whether authors share their code. Consistent rates across both samples therefore suggest that neither filter systematically selects for higher or lower code-sharing behaviour.

\subsubsection{RQ1 -- Code Availability}
Across both datasets, code availability remains limited.
In the manual analysis, 31 out of 127 papers (\ie, \textbf{24.4\%}) provide source code, with an additional 7.9\% offering code only upon request.
This aligns closely with the large-scale analysis, where 26.8\% of papers include accessible repository links and a further 10.8\% mention code availability but do not provide an accessible link, contain explicit \enquote{available upon request} disclaimers instead of public code artefacts, or contain repository links that are no longer reachable.
Taken together, these findings indicate that only about one quarter of papers provide directly usable code artefacts, with a non-negligible fraction relying on restricted or unavailable access mechanisms.

\subsubsection{RQ2 -- Environment Specification}
A similar pattern can be observed for environment specification.
In the manual analysis, 17 out of the 31 papers (\ie, \textbf{54.8\%}) that have code available also provide environment specifications, while an additional 9.7\% of the repositories provide only partial environmental information.
The large-scale analysis yields comparable results, with 54.4\% of the found repositories providing machine-readable environment files, and 11.1\% offering partial specifications (\eg, framework mentions).
This suggests that even if code is available, an environment specification is not consistently provided in a fully reproducible form.

\subsubsection{RQ3 -- Documentation}
The manual analysis showed that \textbf{61.3\%} of all available repositories have some form of documentation, while an additional 25.8\% provide minimal descriptions for the code.
The large-scale analysis confirmed that documentation is mostly consistently available, with even 78.0\% of the considered papers including a README or similar documentation, and 7.1\% having a partial minimal description.
Taken together, these results indicate that documentation is widely available, although its completeness varies.

\subsubsection{RQ4 -- Hardware Specification}
Most of the considered papers that satisfy RQ1 do not rely on real quantum hardware but instead use classical simulations, accounting for 24 out of 31 repositories (\textbf{77.4\%}).

Among the remaining seven papers that both describe the use of quantum hardware and provide code artefacts, only three specify the target hardware in sufficient detail, two provide only partial specifications, and the remaining two provide no hardware specification.

Although the small sample size limits the strength of conclusions from the manual analysis, the large-scale results indicate a similar overall trend. Conditional on satisfying RQ1, 23.5\% of papers provide a complete hardware specification, 43.5\% provide partial information (\eg, mentioning a platform without calibration details), and the remaining 33.1\% either omit hardware information entirely or do not rely on quantum hardware.

\subsubsection{RQ5 -- Executability}
The manual study revealed that of the examined papers with code availability, only about \textbf{35.5\%} were executable, while the remaining 64.5\% failed to execute due to the following failure modes:
\begin{enumerate}
    \item 
    There is no explicit or correct specification of library dependencies and programming language versions (\eg, Python), resulting in floating dependency drifts.
    \item 
    Required licenses and hardware are either not available or difficult to obtain, which hinders testing the provided code.
    \item The provided code contains hardcoded local file paths, as well as missing path definitions and required files, preventing successful code execution.
    \item Non-trivial API incompatibilities are present, obstructing code execution.
    \item Although the code can be brought to a runnable state, execution remains error-prone and insufficient documentation makes it difficult to verify whether the results are correctly reproduced.
    \item Dockerfiles use \texttt{latest} tags instead of stable pinned versions, leading to unreliable container executions.
\end{enumerate}

While the automated analysis cannot assess the quality, executability or reproducibility of the artefacts, the combination of both analyses suggests that reproduction artefacts are not only scarce, but also often of insufficient quality to enable reproducibility, even in the short term.

\section{Discussion}

Our results reveal a substantial gap between the availability of reproducibility artefacts and their actual executability. While approximately one quarter of the analysed papers provide source code (RQ1), and a subset of these further includes environment specifications (RQ2) and documentation (RQ3), the majority of artefacts (64.5\%) fail to execute successfully in a clean environment (RQ5). This discrepancy suggests that the presence of code, documentation, and environment descriptions alone is not a reliable indicator of reproducibility in practice.

A closer inspection of the observed failure modes indicates that many issues arise from incomplete or implicit assumptions about the execution environment. Common problems include missing or unpinned dependencies, unspecified programming language versions, hardcoded local file paths, and incompatibilities with evolving APIs. Notably, these failures frequently occur even when some form of environment specification or documentation is provided, suggesting that such artefacts often remain incomplete. Taken together, these observations point towards a central challenge: critical aspects of the execution context are not fully captured or communicated, hindering reproducibility.

Several factors may contribute to this phenomenon. From a technical perspective, software is often developed and tested in local environments that implicitly encode assumptions about system configuration, available dependencies, and data layout. These assumptions may not be apparent to the original developers and therefore remain undocumented in shared artefacts. From a process-oriented perspective, it is plausible that time constraints in publication cycles, the absence of widely adopted reproducibility standards, and the lack of structured reproducibility checklists contribute to incomplete artefact descriptions. We emphasise that these factors represent plausible interpretations of our observations rather than quantitatively validated causes.

Our findings further suggest that commonly used approaches to environment management only partially address these challenges. Tools such as dependency files (\eg, \texttt{requirements.txt}), virtual environments, and containerisation frameworks are frequently used to support reproducibility. However, our results indicate that these approaches often fail to capture the full execution context. For instance, dependency specifications may omit system-level requirements, while container-based approaches can suffer from implicit assumptions introduced by base images or the use of non-pinned versions (\eg, \texttt{latest} tags), which aligns with other recent findings on the limited build reproducibility of Docker images~\cite{malka2026docker}. More broadly, no single approach has emerged as a widely adopted standard for fully specifying reproducible environments in \gls{qc} research.

These challenges are particularly pronounced in the context of quantum computing. Compared to other domains, \gls{qc} research is characterised by rapidly evolving software frameworks, frequent API changes, and a strong dependence on specialised hardware or cloud-based platforms~\cite{manuel:22:portability}. Access to quantum hardware may be limited or subject to change, and detailed hardware configurations, including calibration states, are often not fully documented. As a result, \gls{qc} artefacts are especially susceptible to both software and infrastructure-related reproducibility issues. While similar challenges exist in other computational fields, such as machine learning or high-performance computing, the combination of rapid tool evolution and constrained hardware access amplifies these issues in \gls{qc}.

In this context, prior work in other domains has explored declarative approaches to environment specification, such as those enabled by \emph{GNU Guix}~\cite{vallet2022toward, courtes2015reproducible} or \emph{Nix}~\cite{hausch2025improving, dolstra2004nix}. These approaches aim to describe complete execution environments in a machine-readable and reproducible manner, thereby reducing implicit assumptions. By capturing entire dependency graphs, including system-level components, they directly address several of the failure modes observed in this study. While such techniques have shown promising results in areas such as bioinformatics and high-performance computing, their adoption remains limited, and their applicability to \gls{qc} workflows has not yet been systematically explored.

At the same time, declarative approaches introduce practical challenges. In particular, they may interact non-trivially with existing ecosystem-specific tooling, such as Python virtual environments or JavaScript package managers, leading to integration inconsistencies in heterogeneous software stacks. However, similar tensions can be observed in container-based workflows, where multiple dependency management layers must be orchestrated. Furthermore, declarative environment approaches have not yet seen widespread adoption in quantum computing workflows. As a result, while they represent a technically promising direction, further work is required to assess their practical integration into existing research practices.

Overall, our study indicates that reproducibility challenges in quantum computing are not solely a matter of artefact availability, but rather of how completely and explicitly the execution environment is specified. Addressing this gap may require both improved tooling and broader community practices that emphasise explicit, machine-readable, and reproducible environment descriptions.

\section{Recommendations}
Ensuring reproducibility requires making implicit assumptions about the execution environment explicit. Our findings indicate that many failures arise not from the absence of artefacts, but from incomplete or underspecified environments. To address this, authors should provide complete, machine-readable environment specifications that capture all relevant aspects of the execution context, including programming language versions, library dependencies, and system-level requirements. In addition, dependencies should be pinned to stable versions to avoid variability introduced by drifting or floating specifications. Reproducibility can be further improved by offering minimal, well-defined entry points into the intended execution environment, for example through a single command that instantiates a development shell and allows the execution of a representative example with expected behaviour. Importantly, such artefacts should be validated across independent environments, for instance by testing on different machines or by multiple contributors, in order to expose hidden assumptions. Aligning the development and reproduction environments can further reduce discrepancies, as environment specifications are then implicitly maintained as part of the development process.

Beyond software environments, reproducibility in quantum computing is inherently constrained by dependencies on specialised hardware and cloud-based platforms. Access to quantum devices is often limited or changes over time, and cloud APIs and service interfaces may evolve or be deprecated. As a result, even fully specified software environments cannot guarantee long-term reproducibility. To mitigate these challenges, authors should document hardware requirements as precisely as possible and provide alternative execution paths, such as simulator-based fallbacks, where applicable. In addition, experimental results should be preserved at the lowest feasible level of abstraction, including raw measurement data and sufficient metadata to reconstruct full post-processing pipelines. Capturing the chain from raw data to final results improves transparency and enables partial reproduction, even when the original hardware or interfaces are no longer available.

To illustrate these recommendations, we provide a reproduction package that serves as a blueprint for reproducible \gls{qc} artefacts \href{https://github.com/lfd/qce26_claim_against_measurement}{online} with a long-term \href{https://doi.org/10.5281/zenodo.21539924}{Zenodo archive} alternative. The package includes a minimal entry point based on a \texttt{Makefile} and containerised execution via Docker, as well as a fully specified container image archived with a persistent identifier. In addition to enabling direct execution, the environment can be rebuilt from source, ensuring transparency and long-term accessibility. The package further integrates the complete experimental pipeline, allowing the paper, including all figures and results, to be recompiled from the underlying data. As an exploratory extension, the package also includes a declarative environment specification based on a Nix flake-based configuration. While this component is not yet intended as a complete solution, it demonstrates how declarative approaches can be used to make environment assumptions explicit and reproducible.

\section{Conclusion}
Reproducibility remains a key challenge in \gls{qc} research.
In this work, we analysed a large corpus of \gls{qc} papers and found that only around one quarter of publications provide accessible code artefacts (24.4\% in the manual study, 26.8\% in the large-scale analysis). Among these, only a minority include complete environment specifications, and the majority of artefacts (64.5\%) fail to execute successfully in a clean environment. These results highlight a substantial gap between artefact availability and actual executability.

Our findings indicate that these failures are primarily driven by incomplete and implicit assumptions about the execution environment rather than by the absence of artefacts. Even when code and documentation are available, critical details such as dependency versions, system-level requirements, or data layout are often underspecified. This is further compounded by domain-specific characteristics of quantum computing, including rapidly evolving software ecosystems and dependencies on specialised hardware and cloud-based platforms, which introduce additional challenges for consistent and long-term reproducibility.

While the practical recommendations outlined in this paper address common failure modes, they do not fully eliminate the underlying issue of incomplete environment specification. Declarative approaches to environment management represent a promising direction in this context; however, their applicability to \gls{qc} workflows remains largely unexplored. Future work will investigate their potential for improving the robustness and longevity of reproducible research artefacts.

\begin{small}\noindent\textbf{Acknowledgements}
We acknowledge partial support by the German Research Foundation, grant MA 9739/1-1, by the High-Tech Agenda of the Free State of Bavaria, the German Federal Ministry of Research, Technology and Space (BMFTR), funding program ‘Research Program Quantum Systems’, grant number 13N17387, the European Regional Development Fund (ERDF) and the Free State of Bavaria as part of the project AIM-SMEs (Grant No. 2506-014-3.2), co-funded by the European Union. 
\end{small}

\printbibliography
\end{document}